# Coexistence of two order parameters and a pseudogaplike feature in the iron-based superconductor LaFeAsO$_{1-x}$F$_x$

R.S. Gonnelli[1]*, D. Daghero[1], M. Tortello[1], G.A. Ummarino[1], V.A. Stepanov[2], J. S. Kim[3], and R. K. Kremer[3].

[1] Dipartimento di Fisica and CNISM, Politecnico di Torino, corso Duca degli Abruzzi 24, 10129 Torino (TO) – Italy
[2] P.N. Lebedev Physical Institute, Russian Academy of Sciences, Leninskiy Prospekt 53, 119991 Moscow, Russia
[3] Max-Planck-Institut für Festkörperforschung, D-70569 Stuttgart, Germany

The nature and value of the order parameters (OPs) in the superconducting Fe-based oxypnictides REFeAsO$_{1-x}$F$_x$ (RE = rare earth) are a matter of intense debate, also connected to the pairing mechanism which is probably unconventional. Point-contact Andreev-reflection experiments on LaFeAsO$_{1-x}$F$_x$ gave us direct evidence of three energy scales in the superconducting state: a nodeless superconducting OP, $\Delta_1$ = 2.8-4.6 meV, which scales with the local $T_c$ of the contact; a larger unconventional OP that gives conductance peaks at 9.8-12 meV, apparently closes below $T_c$ and decreases on increasing the $T_c$ of the contact; a pseudogaplike feature (i.e. a depression in the conductance around zero bias), that survives in the normal state up to $T^* \sim 140$ K (close to the Néel temperature of the undoped compound), which we associate to antiferromagnetic spin fluctuations (AF SF) coexisting with superconductivity. These findings point toward a complex, unconventional nature of superconductivity in LaFeAsO$_{1-x}$F$_x$.

## 1. INTRODUCTION

The recent discovery of high-temperature superconductivity in the rare-earth iron-based oxide systems REFeAsO$_{1-x}$F$_x$ with $T_c$ ranging from 26 K (RE=La)[1] to 43-55 K (RE= Sm, Nd)[2,3,4] has aroused an extraordinary interest in the scientific community. Some of these materials feature the highest critical temperature known so far with the exception of copper-oxide superconductors, and show a number of intriguing similarities and subtle differences with them. The LaFeAsO$_{1-x}$F$_x$ system, for example, has a quasi-two-dimensional structure which consists of charged (LaO)$^{\delta+}$ layers alternating with (FeAs)$^{\delta-}$ layers – the latter apparently playing the key role for the occurrence of superconductivity – and a quasi-2D Fermi surface made up of multiple disconnected sheets, which immediately suggests the possibility of multiple OPs as in MgB$_2$. The parent (undoped) compound LaFeAsO features a long-range antiferromagnetic spin-density-wave order[5, 6] and, on doping with electrons, the magnetic order is disrupted while superconductivity emerges – with $T_c$ up to 26 K according to the original report[1], further increased by high-pressure synthesis[7] or oxygen deficiencies[8]. Moreover, the standard electron-phonon mechanism for the superconducting pairing cannot explain the observed high critical temperature[9]. An unconventional, spin-fluctuation-mediated pairing has been proposed instead[10].

Direct measurements of the superconducting OP are of crucial importance in this context. A direct knowledge of the OP, of its dependence on temperature and of its symmetry in **k**-space, is indeed necessary to clarify the pairing mechanism in a superconductor[11, 12]. The use of spectroscopic techniques like Scanning Tunnelling Microscopy (STM) or Angle-Resolved Photoemission Spectroscopy (ARPES) is unfortunately still hindered by the polycrystalline nature of the samples of this Fe-As superconductor. Point-contact spectroscopy (PCS), instead, allows the determination of the OP also in present-day polycrystalline samples. Unfortunately, PCS measurements recently carried out in oxypnictides[13-17] have led to often contradictory results. For example, early experiments in SmFeAsO$_{1-x}$F$_x$ strongly point toward a single BCS order parameter[13] but other authors claim evidence for multiple $d$-wave OPs in the same material[14]. More recently, we have found evidence for multiple isotropic gaps in this compound[18]. In the case of LaFeAsO$_{1-x}$F$_x$, early PCS data[15] were interpreted as giving evidence of a single $d$-wave order parameter.

Here, we report on PCS measurements in the Andreev-reflection regime that reveal the presence of two order parameters and a pseudogap-like feature in the superconducting state of LaFeAsO$_{1-x}$F$_x$ (La-1111). The smaller OP, $\Delta_1$, is certainly a superconducting one and does not show node lines, but features a non-conventional behaviour as a function of temperature. The second OP is much larger than $\Delta_1$, features a more conventional temperature dependence but apparently closes before $T_c$. As a function of the critical temperature, the first OP is always close to the standard s-wave BCS value, while the second rapidly decreases on increasing $T_c$ and apparently vanishes or merges with the first at $T_c \approx 31$ K. Finally, we reproducibly observed a pseudogap-like depression in the conductance around zero bias, both in the superconducting state and in the normal state, which persists up to temperatures close to the Néel temperature of the parent compound.

## 2. EXPERIMENTAL DETAILS

The polycrystalline samples of LaFeAsO$_{1-x}$F$_x$ with nominal F content x=0.1 were grown according to Ref.19, by solid-state reaction using LaAs, Fe$_2$O$_3$, Fe, and LaF$_3$ as starting materials. A mixture of the four components was ground and cold-pressed into pellets, placed into a Ta crucible, sealed in a quartz tube under argon atmosphere, and annealed for 50 h at a temperature of 1150°C. The samples present large (5-20 μm) crystallites in a more disordered matrix. Resistivity measurements show bulk superconductivity, with a critical temperature (defined at 90% of the resistive transition) $T_c^{on}$=27 K, and a transition width $\Delta T_c$ = T(90%) – T(10%) $\approx$ 4 K. The resistivity actually deviates from its low-temperature linear behaviour already at $T \approx 31$ K, as shown in Figure 1. The local F content, as measured by micro-EDX (energy-dispersive X-ray spectroscopy) is uniform within the crystallites but can vary from one to another (with a spread $\Delta x = 0.02$). This will be reflected in a variation of the local critical temperature of the point contacts and of the order parameters.

Point-contact Andreev-reflection measurements were performed by using a slightly modified technique in which the point contact is not made by pressing a sharp metallic tip against the sample surface, but by putting a small spot ($\varnothing < 50$ $\mu$m) of conducting paste (containing Ag grains whose size is 2-10 $\mu$m ) on the freshly exposed surface of the sample[20, 21]. This ensures a very good thermal and mechanical stability of the contacts, prevents thermal drifts (so that the contact occurs in the same place at any temperature) and, in particular, avoids possible lattice distortions due to the pressure applied by the tip. This is particularly important in these materials, where a systematic effect of the pressure was often seen to drive the emergence of zero-bias anomalies[13,16].

When performing PCS measurements, one should in principle systematically check that every junction fulfills the condition for ballistic transport in the contact – i.e. that the contact size $a$ is smaller than the electronic mean free path $\ell$. In our case, the real contact occurs on a nanometric scale between single Ag grains in the paste and the sample surface. Due to the way the contact is fabricated, the presence of parallel contacts is also possible within the area of the Ag spot [20, 21]. As a consequence, there is no directly accessible information on the size $a$ of the single nano-contacts. Neither the mean free path $\ell$ in this material is well known, nor its extraction from the resistivity by using a simple free-electron model is a very rough approximation because of the small density of charge carriers [22]. This problem is common to all the PCS measurements in oxypnictides appeared so far. In these conditions, the reliability of the measurement can only be judged *a posteriori* by looking at the shape of the conductance curves. When they do not show sharp dips[23] or other signs of heating effects, a ballistic conduction can be assumed and the curves can be used for further analyses.

## 3. EXPERIMENTAL RESULTS

Figure 2 reports three representative examples of raw point-contact conductance curves at various temperatures, up to the normal state (bottom thick line). The voltage is defined to be negative when electrons are injected into the superconductor. Some interesting pieces of information can be gathered by simply looking at these curves.

First, all the curves show two clear peaks at low bias, plus additional structures (peaks or shoulders) at a higher voltage, but no zero-bias peak (ZBP). This is common to all the spectra we measured. The absence of a ZBP in our curves, instead observed (especially in low-resistance contacts) by PCS measurements in this and other Fe-based superconductors[13-17] is probably related to the "soft" point-contact technique we used, which does not involve pressure applied to the sample, and suggests that the ZBP may not be an intrinsic feature of PCS spectra in LaFeAsO$_{1-x}$F$_x$. This clearly excludes the possibility of a *d*-wave symmetry for the superconducting gap [24].

The second thing one can see in Figure 2 is that the normal state measured at $T_c$ (bottom thick line) is always concave, with a systematic asymmetry. This asymmetry is common to all point-contact [13, 14, 16] and tunnel spectra[26,27] measured in Fe-As-based superconductors. As for the concavity, it suggests the existence of a partial suppression in the density of states about the Fermi level in the normal state. In the following, we will refer to this structure as the "pseudogaplike feature", meaning that there is no direct connection with the peculiar phenomenology of the pseudogap in cuprates. A closer inspection of the "tails" of the curves in Fig.2 shows that this pseudogaplike feature is also present below $T_c$ (i.e. it coexists with superconductivity) and becomes deeper and deeper on cooling. Unfortunately, due to the very high critical fields[28] in LaFeAsO$_{1-x}$F$_x$, the actual normal state at low temperature (for example 4.2 K) is not accessible experimentally.

The third thing to notice in Figure 2 is that the shape of the low-temperature curves, with the structures indicated by arrows, suggests the presence of multiple OPs – as a matter of fact, these curves look very similar to those measured in pure and doped MgB$_2$. The possibility of multiple OPs has been proposed, for this material, on the basis of critical-field measurements [28], and it could well be compatible with the Fermi surface, made up of separate sheets[22]: recent angle-resolved photoemission spectroscopy (ARPES) measurements in Ba$_{0.6}$K$_{0.4}$Fe$_2$As$_2$ have indeed shown the opening of different OPs on the various sheets of the Fermi surface, which might be a general property of the Fe-As-based superconductors[29,30].

Let us then assume that the features indicated by arrows in Figure 2 are the hallmark of two superconducting OPs (an alternative analysis of the data, in the hypothesis that only the smaller one is a superconducting OP, is given in Appendix C). The amplitudes of these OPs can then be obtained, as in MgB$_2$, through a fit of the experimental data with a suitable two-channel model in which the normalized point-contact conductance $G$ is the weighted sum of two Blonder-Tinkham-Klapwijk (BTK) contributions[31] generalized to take into account broadening effects [20,21,32] and the angular distribution of the injected current at the interface[25]: $G = w_1 G_1^{BTK} + (1-w_1) G_2^{BTK}$. The parameters of this model – that we will call "generalized two-band BTK model"– are the OP magnitudes $\Delta_1$ and $\Delta_2$, the potential barrier parameters $Z_1$ and $Z_2$, the broadening parameters $\Gamma_1$ and $\Gamma_2$, plus the weight $w_1$ (see Appendix B for details). Prior to this fitting procedure, a normalization (i.e. division by the "normal-state" conductance, when $\Delta_1 = \Delta_2 = 0$) must be performed. As noted above, here the normal-state conductance at $T < T_c$ is certainly more concave than that measured at $T_c$ but is not directly measurable. We therefore tried to simulate it by using a B-spline curve that connects the tails of the conductance curves (at V>20 mV) with a suitable point at zero bias. Further details are given in Appendix A, where we also compare this procedure with a second kind of normalization that consists in simply dividing all the conductance curves by the experimental normal state at $T_c$, vertically translated if necessary. Although both procedures are necessarily approximated and contain some elements of arbitrariness, it will be shown that the final results (in particular, the trend of the OPs as a function of temperature) are rather robust against the choice of the normalization process.

Figure 3a shows an example of conductance curves normalized with the first method (symbols) compared to the relevant two-gap fit (solid lines) as a function of temperature (the raw data are those shown in Figure 2a); the inset shows a comparison between the single-gap fit (dashed line) and the two-gap one (solid line). The OP values resulting from the two-gap fit of the curves shown in Fig.3a are reported in Fig.3b as solid symbols; triangles in the inset represent the broadening parameters $\Gamma_1$ and $\Gamma_2$, while the temperature-independent values of the barrier parameters $Z_1$ and $Z_2$ and of the weight $w_1$ are indicated in the label. The gap ratios are $2\Delta_1(0)/k_BT_c= 3.23\pm0.34$ and $2\Delta_2(0)/k_BT_c= 8.5\pm0.5$, in good agreement with the recent findings of NQR in LaFeAsO$_{0.92}$F$_{0.08}$[35]. The decrease of the OPs on increasing the temperature, although regular and smooth, is definitely non-conventional. The larger OP, $\Delta_2$, actually shows a roughly BCS-like behaviour up to about 22 K but, above this temperature, it either disappears or becomes so small that it is impossible for the measurement to discern it. The smaller OP, $\Delta_1$, follows a non-BCS temperature dependence with a "tail" up to the critical temperature of the contact, when the normal-state conductance is recovered. Note that it is very unlikely that this "tail" belongs to the large gap, since the last few conductance curves below $T_c$ present a small peak at zero bias that absolutely cannot be reproduced with the values of $\Gamma_2$ and $Z_2$ – unless one admits an abrupt discontinuity in these parameters. The odd behaviour of $\Delta_1$ and $\Delta_2$ has been observed in all the contacts we measured and it can be easily shown not to depend on either the normalization or the fitting procedure. As a matter of fact, the choice of the normalization affects the values of the OPs (especially $\Delta_2$), but not their trend as a function of temperature, as shown in Fig.3b (open symbols). Moreover, the unconventional "tail" of $\Delta_1$ does not depend on whether we perform a two-band BTK fit of the whole conductance curves *or a single-band BTK fit* of their central part, disregarding the structures related to $\Delta_2$ as in Ref. 13 (see Appendix C for details). Finally, it is worth noting that the unconventional temperature dependence[34] of $\Delta_1$ and $\Delta_2$ is not an artefact due to the point-contact technique we used, because the same measurements carried out in other systems like MgB$_2$ (Ref.20) and CaC$_6$ (Ref. 21) gave temperature dependencies of the OPs in almost perfect agreement with theoretical predictions, with no high-temperature anomalies.

As a further example, Figure 4a reports the conductance curves of Fig. 2c after normalization (with the first method). The experimental data (symbols) are compared to the relevant generalized two-band BTK fit (solid lines). In the inset the two-gap fit (solid line) of the low-temperature conductance curve (symbols) is compared to the single-gap one (dashed line). The OPs resulting from the two-band fit are shown, together with the other fitting parameters, in Fig 4b. The similarity with Fig. 3b is clear although these two figures refer to two contacts with different resistance, placed in a different position on the sample surface, and featuring a different local critical temperature.

For the same contact as in Fig.4a, we also studied the effect of magnetic fields on the conductance curves. Figure 5a reports the experimental normalized conductance (symbols) measured at 1.8 K in magnetic fields of increasing intensity up to 8 T, compared to the relevant generalized two-band BTK fit (lines). The OPs extracted from the fit are reported in Figure 5b as a function of the magnetic field intensity (solid symbols). The slope of the two trends (roughly linear, on this scale) is rather different and might indicate that $\Delta_2$ is more sensitive to the magnetic field than $\Delta_1$.

As already pointed out, the different local F content in the crystallites gives rise to a distribution in local $T_c$ values (already evident in Figure 2) but also affects the OP values. Figure 6a reports the normalized conductance curves of a 15-$\Omega$ contact with $T_c$ =31.0 K. This is an unusually high value for the critical temperature in LaFeAsO$_{1-x}$F$_x$, although bulk $T_c$ values higher than 26 K have been reported for samples synthesized at high pressure[7] (up to 41 K) or with oxygen deficiencies[8] (31.2 K). The latter possibility cannot be excluded in our samples, so that F doping and O deficiencies might concur in determining the local $T_c$ of the crystallites. It is interesting to note that this very high local critical temperature corresponds to the deviation of the $\rho(T)$ curve from linearity (see Figure1). The experimental curves of Fig 6a (symbols) can be fitted very well with a generalized *single*-band BTK model (with only $\Delta$, $\Gamma$ and $Z$ as parameters). The fitting curves are shown as lines in Fig. 6a and the best-fitting values of the order parameter $\Delta$ are reported in Fig. 6c. In this case the use of the alternative normalization (division by the shifted normal state at $T_c$) gives $\Delta$ values that, at any temperature, differ by less than 10%. The ratio $2\Delta(0)/k_BT_c$ is equal to 3.44 – close to the standard BCS value for weak coupling in s-wave – but the temperature dependence is non-BCS and features the characteristic tail already observed in Figure 3b. The persistence of this anomaly also in the absence of a larger OP automatically excludes any possibility of proximity effects due to $\Delta_2$ as the origin of the tail (on the other hand, this is already ruled out by the apparent closing of $\Delta_2$ before $T_c$). The presence of a single OP is also shown by the magnetic-field dependence of the conductance curves (Figure 6b) and by the corresponding plot of $\Delta$ as a function of the field (Figure 6d). A fit of these points (solid line) with a function $\Delta(B)=\Delta(0)\sqrt{1-B/B_{c2}}$ (as expected for a bulk type-II superconductor [11,12]) would lead to B$_{c2} \approx 55$ T, in reasonable agreement with direct critical-field measurements [28]. However, it is worth noticing that a two-gap fit is possible as well and that the presence of $\Delta_2$ cannot be completely excluded. This is shown in the inset of Figure 6a, where the two-gap fit (solid line) of the low-temperature conductance curve (symbols) is compared to the single-gap one (dashed line). Although the single-gap fit is slightly better, the two fits look comparable. Moreover, in the two-gap fit, $\Delta_2$ ($\approx 5.4$ meV) is very close to $\Delta_1$ ($\approx 4.4$ meV) and their error bars overlap, which means that the two gaps might not be distinguishable (see Fig. 7).

The values of $\Delta_1$ and $\Delta_2$ obtained in contacts with different local $T_c$ values are reported as a function of the $T_c$ itself in Fig. 7 (solid and open circles) together with the gap value obtained from the one-gap fit (open squares). Although at present the number of measurements is small, a picture seems to emerge in which $\Delta_1$ slightly increases with $T_c$ while $\Delta_2$ quickly decreases, until a single (or two, very close to each other), superconducting order parameter with BCS value is observed in the highest-$T_c$ contact. If the critical temperature is related to the effective local F content, Figure 7 can be converted in a phase diagram for the LaFeAsO$_{1-x}$F$_x$ compound. This result needs verification from further measurements in samples with different, well-controlled doping content across the superconducting dome. If confirmed, it would largely enrich our knowledge of the phase diagram for LaFeAsO$_{1-x}$F$_x$ and establish interesting similarities between the behaviour of $\Delta_2$ and that of the pseudogap in high-T$_c$ cuprates [35] (note that, however, here $\Delta_2$ is *not* a pseudogap).

Let us then investigate the evolution of the normal-state conductance with temperature. To do so, we measured the conductance curves of our point contacts up to T $\approx$ 200 K, also expanding the voltage range up to about 100 mV. The results for two contacts are shown in Figure 8; note that panel b refers to the contact with the highest T$_c$. The normal state at T$_c$ (upper thick line) is asymmetric and "pseudogapped", with two broad structures (at energies of the order of 50 meV, thus much greater than both $\Delta_1$ and $\Delta_2$) that are progressively smoothed out on increasing the temperature. Their shape is very similar to that observed by PCS in materials with long-range spin-density-wave (SDW) order, like URu$_2$Si$_2$ [36, 37]. No such long-range SDW order was observed in superconducting LaFeAsO$_{1-x}$F$_x$ by neutron scattering or muon spin relaxation studies [38, 39]; however, these techniques can only detect quasi-static magnetic order while photoemission and tunnelling (or Andreev-reflection) spectroscopies, being sensitive to electron dynamics, can detect phenomena on a much shorter time scale. The proximity of the superconducting state to the antiferromagnetic (AF) SDW state of the parent compound [5, 6] indeed suggests the persistence, in the doped compound, of spin fluctuations that have already been invoked as the origin of the pseudogaplike feature, also observed by photoemission spectroscopy [40, 41] in the same material. This picture has been recently substantiated theoretically by showing that, in systems with reduced dimensionality (< 3D), local AF fluctuations, even on a short range, can give rise to corrections in the self-energy of quasiparticles that, in turn, allow a pseudogap to be opened at the Fermi level [42]. The connection between the pseudogaplike feature of our curves and spin fluctuations is further supported by the fact that the pseudogaplike feature is progressively filled on increasing the temperature and completely disappears at T* $\sim$ 140 K (see Figures 8a and 8b), even in the contact with the highest T$_c$. This temperature is strikingly close to the temperature at which long-range magnetic order appears in the parent, undoped compound [5, 6].

## 4.CONCLUSIONS

To summarize, point-contact spectroscopy measurements in the new Fe-As based superconductor LaFeAsO$_{1-x}$F$_x$ gave us the first evidence of three energy scales coexisting in the superconducting state of this novel superconductor. The first is a nodeless superconducting gap $\Delta_1$ that gives rise to clear coherence peaks in the conductance curves, is almost BCS in value, scales with T$_c$ but displays a non-BCS temperature dependence with a high-temperature "tail". The absence of node lines crossing the Fermi level is suggested by the absence of zero-bias conductance peaks in the conductance curves. The high-temperature "tail" is present in all the contacts, and does not depend on either the normalization of the curves or the model used to fit them (single-band BTK for the fit of the central part alone, as in Ref. 13, or two-band BTK for the fit of the whole conductance curve). Incidentally, this "tail" is very similar to that observed by PCS in the heavy-fermion superconductor CeCoIn$_5$ (Ref. 43).

The second energy scale is related to a larger order parameter, that gives rise to peaks or shoulders in the conductance curves, well distinct from those related to $\Delta_1$. If this second order parameter is interpreted as being a superconducting one, its amplitude $\Delta_2$ can be extracted from a generalized two-band BTK fit of the conductance curves. $\Delta_2$ turns out to have a more conventional temperature dependence than $\Delta_1$ but seems to close at T < T$_c$, has a very high ratio 2$\Delta_2$(0)/k$_B$T$_c$ >5, and decreases on increasing the critical temperature of the contact (in a striking similarity with cuprates) until it disappears, or becomes almost coincident with ₁ when T$_c$=31 K. All these features are unconventional, in the sense that they are inexplicable within a standard two-band Eliashberg theory for boson-mediated superconductivity [44]. This might suggest for the large OP a non-superconducting (maybe magnetic) origin; in this case its amplitude can only be estimated by looking at the position $\Delta_{peak}$ of the peaks/shoulders in the conductance curves. It can be shown (see Appendix C) that even in this case the trend of $\Delta_{peak}$ as a function of the temperature, of the magnetic field and of the critical temperature of the contact would remain qualitatively the same.

The third energy scale is on the order of 50 meV or more and is set by a pseudogaplike depression in the PCS spectra that, although clearly visible at T $\geq$ T$_c$, is also present in the superconducting state and influences the shape of the superconducting conductance curves. This feature turns out to be progressively filled on increasing the temperature but persists well above T$_c$ and up to the Néel temperature of the undoped compound (T*$\sim$140 K). This finding points towards a relationship between AF spin fluctuations (which remind of the long range antiferromagnetic order in the undoped compound) and the pseudogaplike feature, and then suggest that AF SF coexist with superconductivity below T$_c$.

Let us conclude by saying that these results, if compared to recent findings of PCS in other Fe-As superconductors of the 1111 family [SmFeAsO$_{1-x}$F$_x$ (Ref. 18), NdFeAsO$_{1-x}$F$_x$ (Ref. 17)] and of the 122 family [Ba$_{0.55}$K$_{0.45}$Fe$_2$As$_2$ (Ref.45)] as well as with ARPES results in Ba$_{0.6}$K$_{0.4}$Fe$_2$As$_2$ (Refs. 29,30), point toward a general picture for Fe-As based superconductors despite the differences between the various materials. Within this unified picture, two OPs with different amplitude open up on different sheets of the Fermi surface, the smaller more or less BCS in value, and the larger with a strikingly non-BCS amplitude. Moreover, a pseudogaplike feature, probably related to spin fluctuations, persists in the normal state. If confirmed, these signs of convergence towards a unified picture of Fe-As based superconductors would be an important result and would pose serious constraints on any theoretical model for superconductivity in these compounds.

## ACKNOWLEDGMENTS

The authors wish to thank L. Boeri, R. De Renzi, O. Dolgov, S. Massidda, G. Sangiovanni, A. Toschi, for enlightening discussions. Special thanks to I.I. Mazin for kind advice, important comments and continuous encouragement. This work was partially supported by the PRIN project No. 2006021741. V.A.S. acknowledges support by the Russian Foundation for Basic Research Project No. 09-02-00205.

## APPENDIX A – NORMALIZATION OF THE CONDUCTANCE CURVES

In this section we give further details about the normalization of the raw conductance curves, which is necessary to extract the OP values through a generalized two-band BTK fit. As previously said, the normalization is not straightforward since the shape of the normal state depends on temperature – the zero-bias dip, or pseudogaplike feature, being progressively filled on increasing temperature. This is evident above $T_c$, but can be inferred also in the superconducting state by looking at the tails of the conductance curves.

In panel (a) of Figure A1 we report a subset of the conductance curves shown in Fig. 2a. By comparing the normal-state conductance at $T_c$=28.6 K (open circles) to the curve at T=21.8 K (light grey circles) the progressive "filling" of the background (and the consequent increase of normal-state zero-bias conductance, NZBC) on increasing the temperature is clear. Therefore the normal state at T=4.3 K should coincide with that at $T_c$ for V>20 mV, but should also feature a smaller NZBC. The solid line, constructed by connecting the black points with a B-spline function, is a guess for that normal state and is the curve by which the experimental data at T=4.3 K (dark grey circles) were divided. The same procedure was used for the other temperatures, respecting the progressive increase in the NZBC. The results of the fit performed after this normalization were shown in Fig. 3b and, for other point contacts, in Fig. 4b, 5b and 6c.

In panel (b) of Figure A1, a different possible choice for the normalization is described. Here we simply used the normal-state conductance at $T_c$ (vertically translated if necessary) to normalize all the curves. For example, the curve at T=4.3 K was divided by the normal state at $T_c$ (solid line) while the curve at T=21.8 K was divided by the normal state translated downward of a suitable amount (dashed line). This second procedure clearly gives rise to some anomalies at high bias, where the normalized curves will depart from unity (the same happens in Ref. 13, Supplementary Information). The amplitudes of $\Delta_1$ and $\Delta_2$ obtained from the fit of the conductance curves normalized in this way were reported in Figure 3b as open symbols. Despite the large difference in the normalization, at low temperature the amplitude of $\Delta_2$ changes by less than 10% and that of $\Delta_1$ by less than 1.5%.

## APPENDIX B – THE GENERALIZED TWO-BAND BTK MODEL

The fit of the normalized curves, in the hypothesis that both the order parameters are superconducting, was performed by supposing that the conductance can be expressed as the sum of suitably weighted contributions

$$G(E) = w_1 G_1^{BTK}(E) + (1-w_1)G_2^{BTK}(E)$$

where each conductance is given (at T=0, for simplicity) by the Blonder-Tinkham Klapwijk model [31] generalized to the 3D case [25]:

$$G_i^{BTK}(E) = \frac{\int_{-\pi/2}^{\pi/2} \sigma_{S,i}(E,\phi)\cos(\phi)d\phi}{\int_{-\pi/2}^{\pi/2} \sigma_{N,i}(\phi)\cos(\phi)d\phi}$$

where $i=1,2$ is the band index, $\phi$ is the angle the direction of injection makes with the normal to the interface, and

$$\sigma_{N,i}(\phi) = \frac{\cos(\phi)^2}{\cos(\phi)^2 + Z_i^2}$$

$$\sigma_{S,i}(E,\phi) = \sigma_{N,i}(\phi) \frac{1 + \sigma_{N,i}(\phi)|F_i(E)|^2 + (\sigma_{N,i}(\phi)-1)|F_i(E)^2|^2}{\left|1 + (\sigma_{N,i}(\phi)-1)F_i(E)^2\right|^2}$$

where the parameters $Z_1$ and $Z_2$ are related to the height of the potential barrier at the interface. The functions $F_i(E)$ are given by

$$F_i(E) = \frac{(E + i\Gamma_i) - \sqrt{(E+i\Gamma_i)^2 - \Delta_i^2}}{|\Delta_i|}$$

and contain the order parameters $\Delta_1$ and $\Delta_2$. As a further generalization, we included in the model the broadening parameters $\Gamma_1$ and $\Gamma_2$ as imaginary parts of the energy E [32]. In our case, $\Gamma_1$ and $\Gamma_2$ account for both the (small) intrinsic lifetime broadening and other effects – related to the experimental technique and thus extrinsic – that smooth the curves and decrease the amplitude of the Andreev signal. Their values are always smaller than the corresponding OP amplitude, and increase on increasing temperature and magnetic field intensity. In the absence of theoretical indications, the weight $w_1$ was taken as an adjustable parameter to be fixed by the fit at the lowest T and B=0 and, then, to be kept constant in all the fits at higher temperature and magnetic field. Actually, in most cases we got $w_1$=0.5 - 0.6. The parameters $Z_1$ and $Z_2$ that enter equation for $\sigma_{N,i}(\phi)$ are related to the height of the potential barrier at the interface and are independent of both temperature and magnetic field.

## APPENDIX C – ANALYSIS OF THE PCS SPECTRA IN THE CASE OF A NON-SUPERCONDUCTING ORIGIN OF THE LARGER OP.

One important question arising from the results shown in Figs. 2,3,4 and 6 is whether $\Delta_1$ and $\Delta_2$ are both superconducting OPs or, instead, one of them has a different origin. In the main text we assumed that they are both superconducting OPs, which seems to be more and more confirmed by other experimental findings in Fe-As-based superconductors. However, at the present stage of knowledge we cannot completely exclude a different hypothesis. Some experimental facts find a simpler explanation if one admits that $\Delta_2$ is not a superconducting OP: i) the anomalous temperature dependence, with the large OP $\Delta_2$ apparently closing at T<$T_c$; ii) the very high value of the ratio $2\Delta_2(0)/k_BT_c$ (at the lowest $T_c$, on the order of 8); iii) the decrease of $\Delta_2$ on increasing $T_c$; iv) the faster decrease of $\Delta_2$ in magnetic field. The first three findings find no explanation within the standard two-band Eliashberg model [44] even if a weak (very strong) coupling and a very large (small) effective boson frequency are used for the band related to $\Delta_1$ ($\Delta_2$).

If we suppose the large OP not to be superconducting (but, for example, of magnetic origin) we clearly cannot obtain its amplitude by means of the generalized two-band BTK fit. In this case, instead, the normalized point-contact conductance can be expressed as

$$G(V) = w_l G_1^{BTK}(V) + (1 - w_l) G_2^{tunn}(V)$$

which is the weighted sum of a BTK contribution (containing $\Delta_1$, $\Gamma_1$ and $Z_1$), and of a contribution from plain tunnelling into a density of states $N_2(E)$ which is not flat but contains structures determined by the larger OP:

$$G_2^{tunn}(V) = \int_{-\infty}^{\infty} N_{2,N}(E) \left[ -\frac{\partial f(E + eV)}{\partial (eV)} \right] dE$$

Here $N_{2,N}(E)$ is the normalized DOS and $f(E)$ is the Fermi function at the temperature at which the measurement was carried out. In principle, at T=0 the conductance $G_2^{tunn}(V)$ coincides with the normalized DOS, while at T≠0 it contains a thermal broadening term.

The shape of the tunnel-conductance contribution $G_2^{tunn}(V)$ cannot be extracted directly through a fit of the conductance curves since the functional form of $N_{2,N}(E)$ is not known. One could think to simulate it with a smeared BCS-like tunnel spectrum, but: i) this would again implicitly assume that the second OP is superconducting; ii) this is exactly what one does using the two-band BTK model (at least if $Z_2$ is large, as in the case of Fig.3).

The best thing to do is thus to extract $G_2^{tunn}(V)$ from the experimental conductance curve $G^{exp}(V)$ (normalized so as to get rid of the pseudogaplike feature) by subtraction of the superconducting contribution.

For example, let us start from the low-temperature normalized curve of Fig.3a and suppose that it is the superposition of a superconducting contribution (related to the small gap $\Delta_1$) and of a non-superconducting one, for simplicity with equal weights: $G(V) = 0.5\ G_1^{BTK}(V) + 0.5\ G_2^{tunn}(V)$. The numerical factor 0.5 is necessary to ensure a correct normalization. The superconducting part can be approximately determined by fitting only the central part of the curve (as it was done in Ref.13) and the non-superconducting part can be determined by subtraction. The two partial contributions to the conductance, $G_1^{BTK}(V)$ (theoretical curve that fits the central part of the experimental conductance) and $G_2^{tunn}(V)$ (experimental points after subtraction) are depicted in Fig. C1.

The BTK fit of the central part of the curve was obtained by using $\Delta_1$=2.67 meV, $\Gamma_1$=1.07 meV, $Z_1$=0.25. These parameters differ only slightly from those given by the generalized two-band fit of the same curve. The curve shown by symbols in Fig.C1, representing $G_2^{tunn}(V)$, has clear peaks at about ±9.5 mV whose position can be used as an evaluation of the larger OP, if it is believed to be non-superconducting.

The same procedure can be followed at any temperature. The resulting tunnel contributions to the conductance are depicted in Fig. C2a and the energy position of their peaks, $\Delta_{peak}$, are reported in Fig. C2b, together with the values of the small gap $\Delta_1$ used in the fit of $G_1^{BTK}$ (open symbols). The values of the two gaps obtained from the generalized two-band BTK fit of the conductance curves are also reported for comparison (solid symbols). Note that the value of the small gap changes very little and its high-temperature "tail" does not depend on whether the fit is two-band or single-band.

Incidentally, the $G_2^{tunn}$ curves and the temperature dependence of $\Delta_{peak}$ are strikingly similar to the findings of ARPES concerning the largest order parameter in $Ba_{0.6}K_{0.4}Fe_2As_2$[29,30].

All the data presented in the main text of the paper can be re-interpreted by supposing that the larger order parameter is not superconducting, and thus applying the same procedure described above. The values of the amplitude of the large OP are different from those obtained with a two-band BTK fit, but the general trend of this amplitude as a function of the magnetic field and of the critical temperature of the junction is the same. This is clearly shown in Fig. C3(a), which reports the magnetic field dependence of the gaps (solid symbols, obtained from the curves of Fig.5a) compared to the corresponding values of $\Delta_1$ and $\Delta_{peak}$ (open symbols). In Fig. C3(b), the dependence of the OPs on the local critical temperature is finally shown, obtained from the analysis of the low-temperature conductance curves of different contacts.

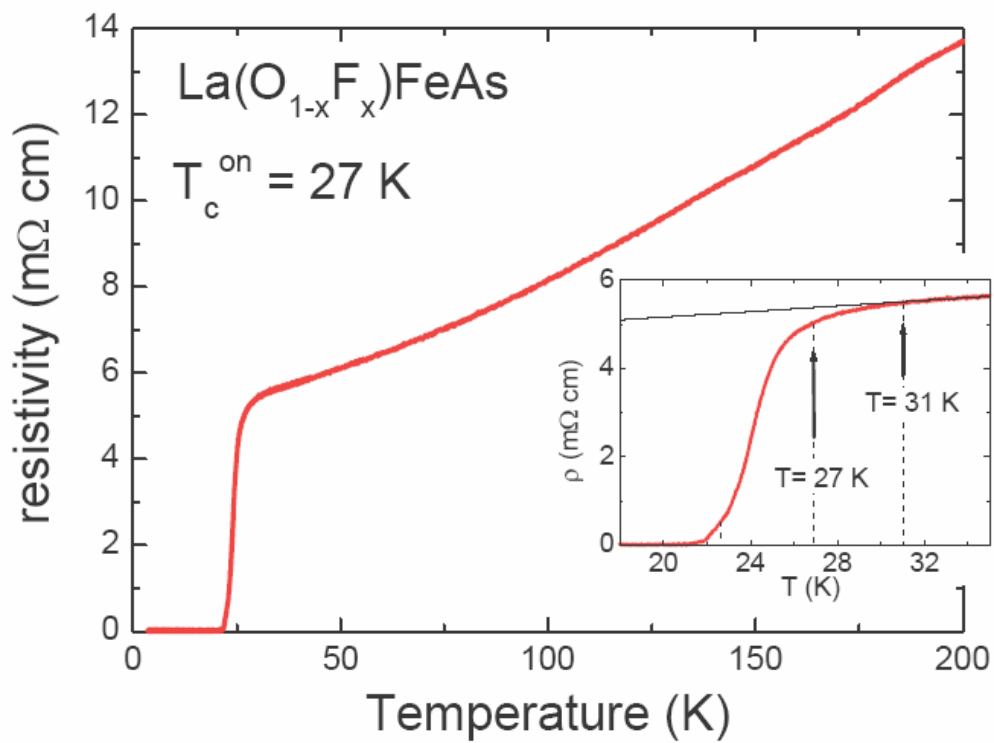

**Figure 1** (color online)

The resistivity of a La-1111 polycrystalline sample used for the PCS measurements. The inset shows a magnification of the superconducting transition. Arrows indicate the deviation from the low-temperature linear behaviour (T=31 K) and the onset of the superconducting transition defined as $T_c^{on}$ = T(90%)=27 K.

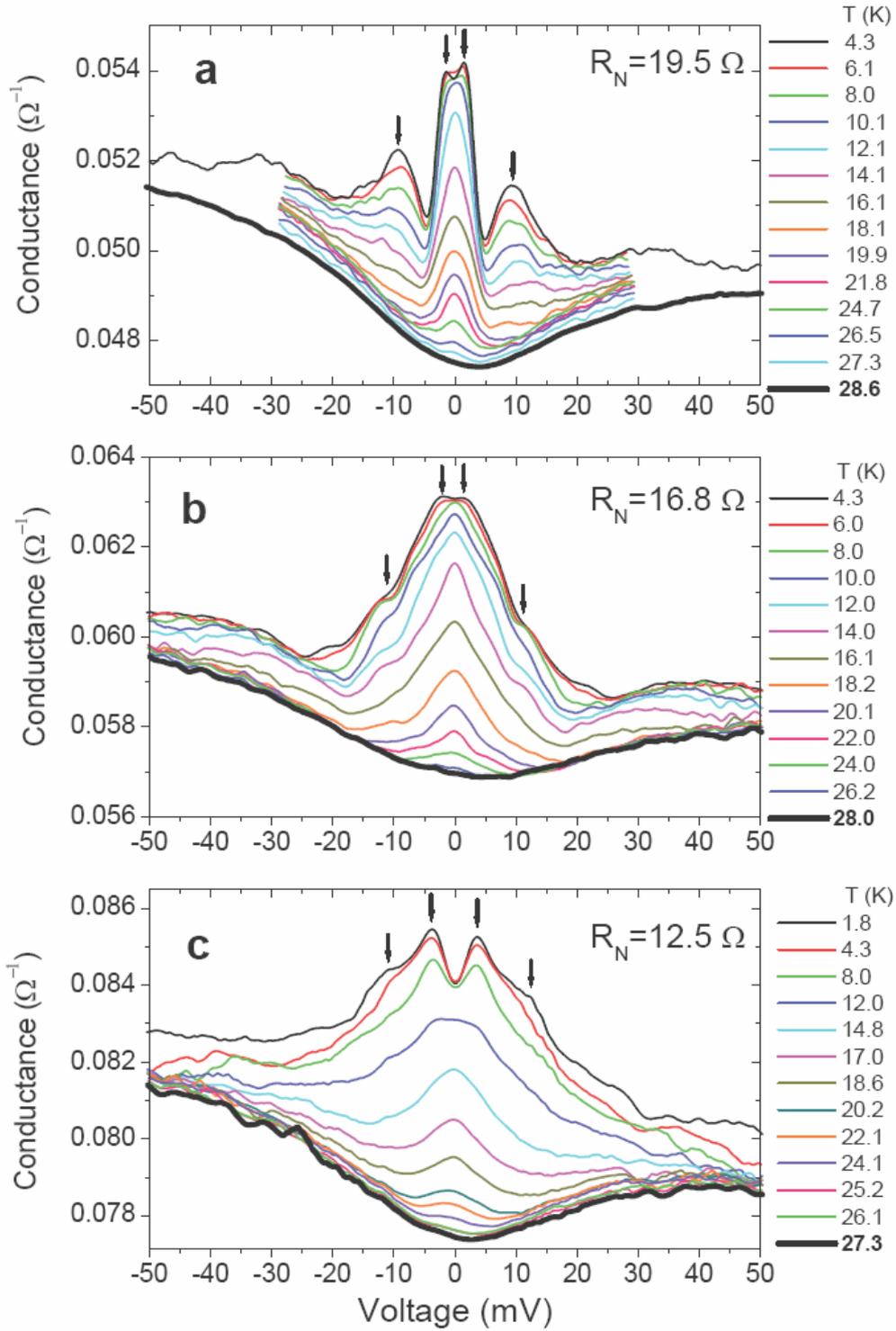

**Figure 2** (color online)

Three examples of raw conductance curves measured in different point contacts on the same polycrystalline LaFeAsO$_{1-x}$F$_x$ sample. The contact resistance R$_N$ is indicated in the label. Arrows indicate the two peaks related to the smaller order parameter $\Delta_1$, and the (more or less pronounced) features related to a second, larger energy scale, $\Delta_2$. The thick black line in each panel represents the normal-state conductance measured at T$_c$.

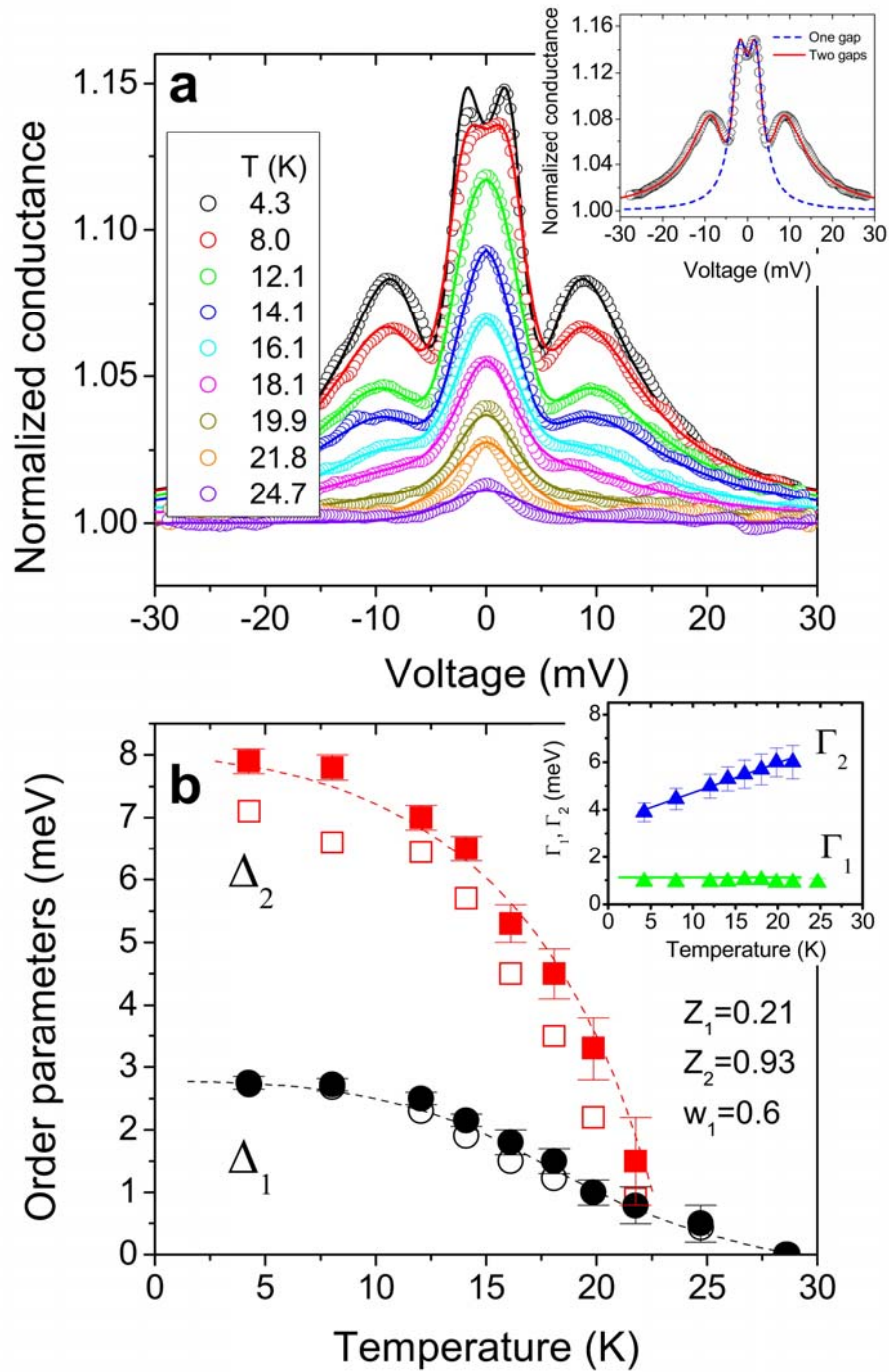

**Figure 3** (color online)

(a) Symbols: conductance curves of a point contact with $T_c$=28.6 K (the same as in Figure 2a) after normalization. The normal state at $T<T_c$, experimentally inaccessible, was here estimated as explained in Appendix A. Solid lines indicate the generalized two-band BTK fit. Inset: comparison between the two-gap (solid line) and single-gap (dashed line) BTK fit of the low-temperature conductance curve (symbols). (b) Temperature dependence of the amplitudes of the OPs, $\Delta_1$ and $\Delta_2$, extracted from the two-band BTK fit (solid symbols). Error bars indicate the uncertainty related to the fitting procedure. Dashed lines are guides to the eye. The inset reports the values of the broadening parameters $\Gamma_1$ and $\Gamma_2$; the values of the barrier parameters $Z_1$ and $Z_2$ and of the weight $w_1$ are temperature-independent and are indicated in the label of the main panel. Open symbols indicate the values of $\Delta_1$ and $\Delta_2$ obtained with a different normalization (see Appendix A for details). The uncertainty (not indicated) is of the same order of magnitude as in the case of solid symbols.

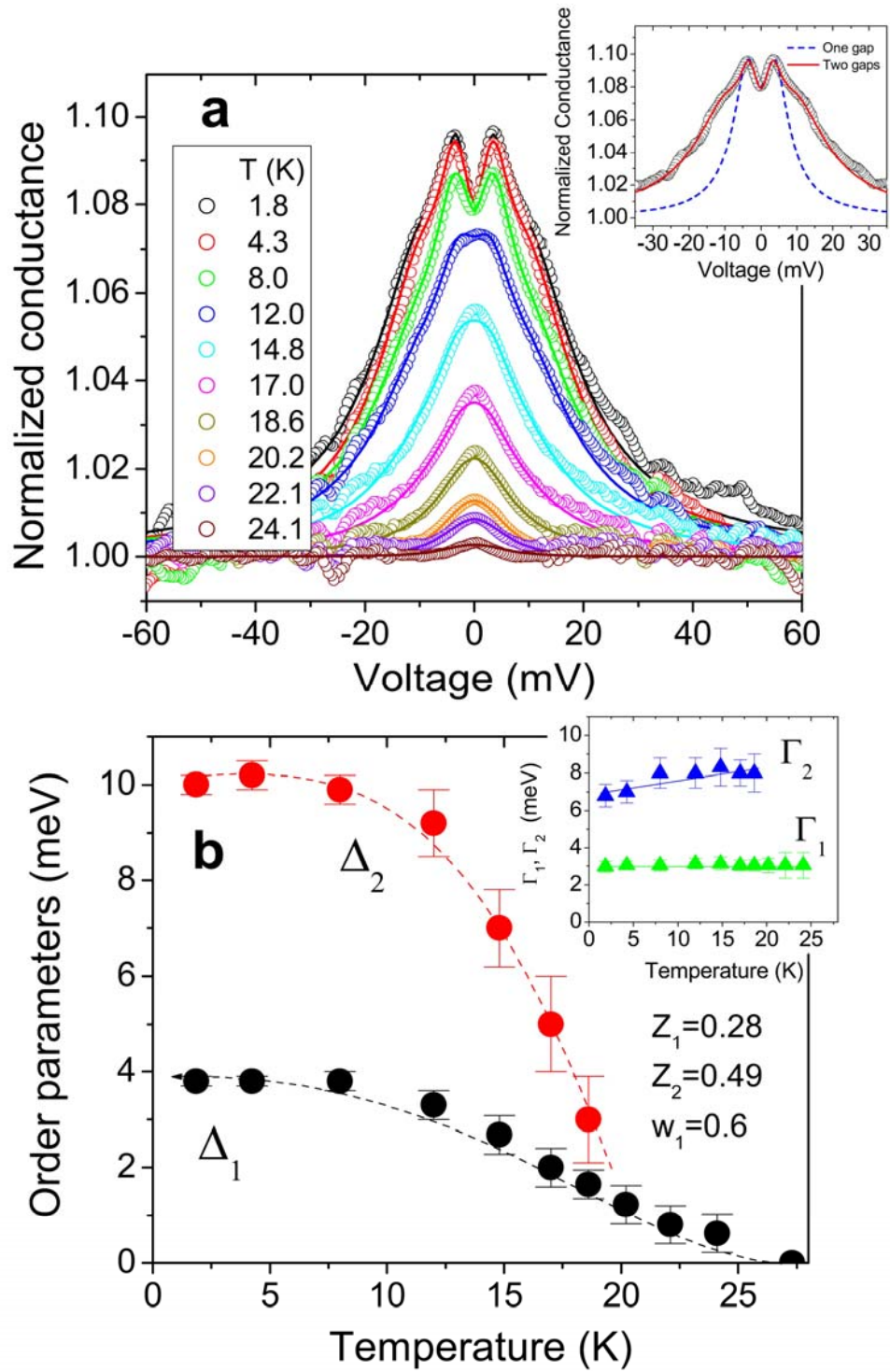

**Figure 4** (color online)
(a) Temperature dependence of the conductance curves of a contact with $T_c$=27.3 K (the same as in Fig.2c). The experimental curves (symbols) were normalized as in Fig.3; solid lines represent the generalized two-band BTK fit. Inset: comparison between the two-gap (solid line) and single-gap (dashed line) BTK fit of the low-temperature conductance curve (symbols). (b) Temperature dependence of $\Delta_1$ and $\Delta_2$ as obtained from the two-band BTK fit. The values of the other fitting parameters are also indicated in the label and in the inset. Lines are guides to the eye. The similarity with Fig. 3b is clear.

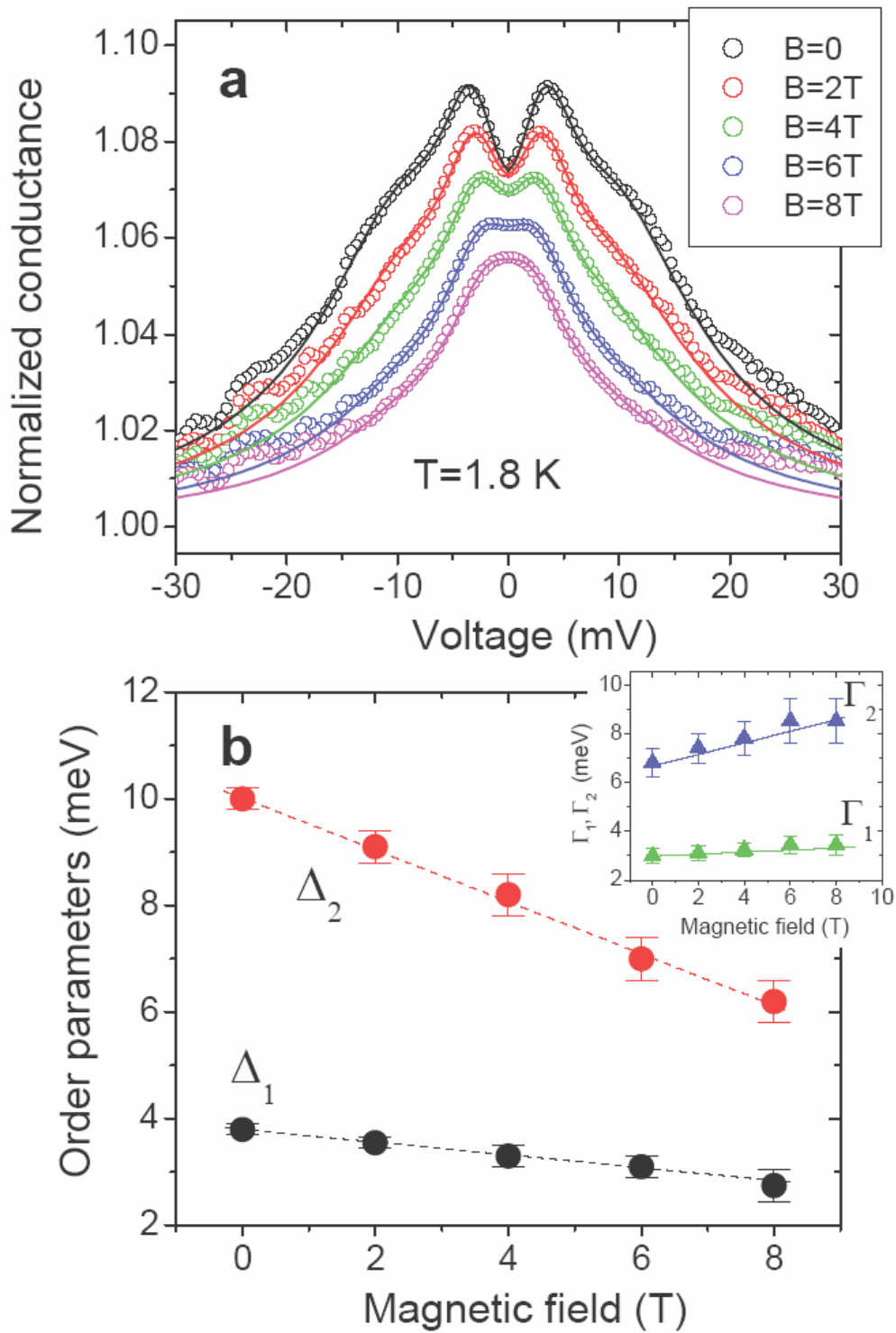

**Figure 5** (color online)
(a) Magnetic-field dependence of the conductance curves of the same point contact as in Fig.4. All the measurements were taken at 1.8 K. Symbols represent experimental data, lines are the best-fitting curves. (b) Magnetic-field dependence of $\Delta_1$ and $\Delta_2$ extracted from the fit of the curves shown in (a). Lines are only guides to the eye. The inset reports the magnetic-field dependence of the broadening parameters $\Gamma_1$ and $\Gamma_2$, while $Z_1$, $Z_2$ and $w_1$ are the same as in Fig.4.

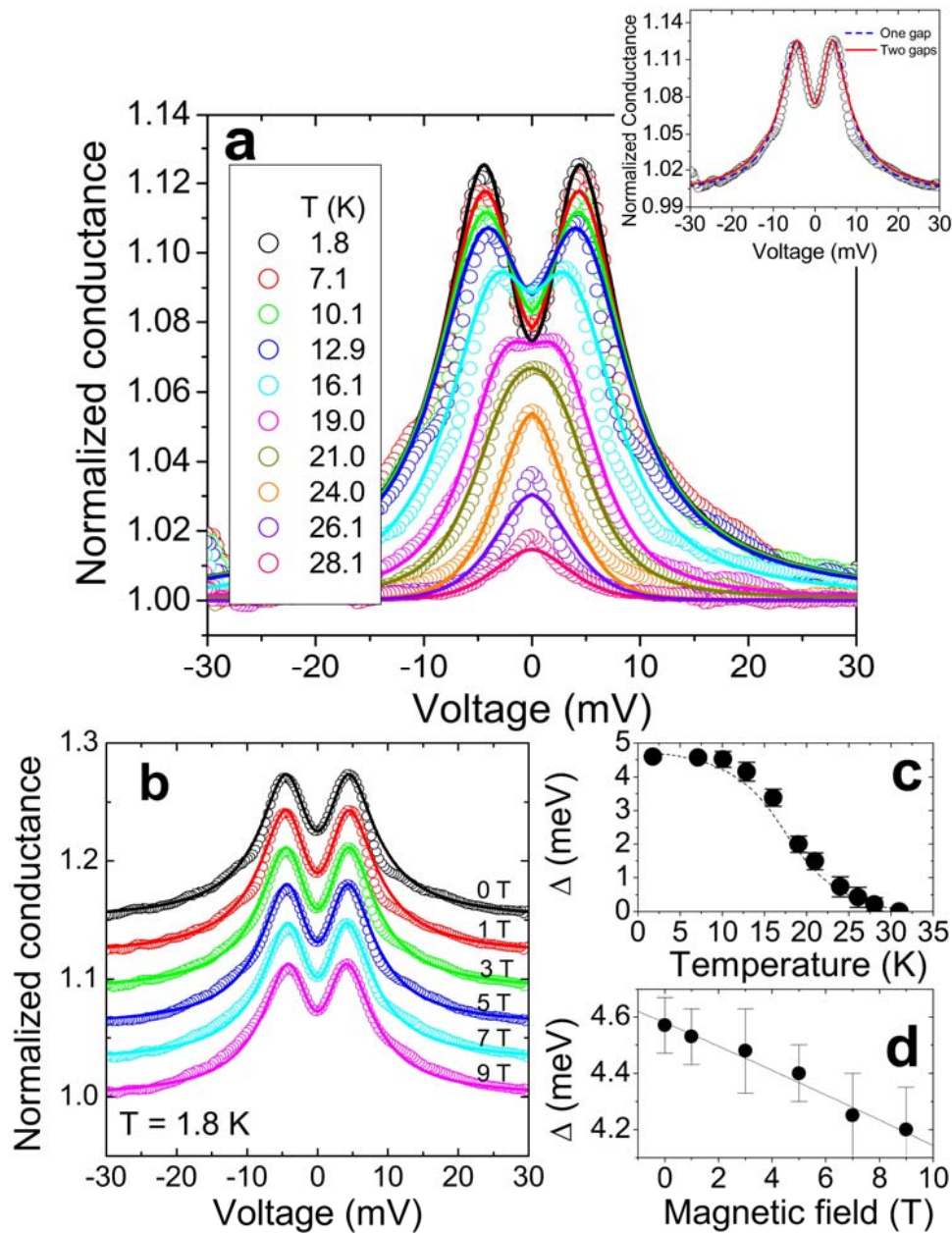

**Figure 6** (color online)
(a) Conductance curves of a 15-Ω point contact with $T_c$=31.0 K after normalization (symbols) together with the generalized single-band BTK fit (lines). At T=1.8 K the best-fitting parameters are: Δ= 4.59 meV, Γ=3.27 meV, and Z=0.40. Inset: comparison between the two-gap (solid line) and the single-gap (dashed line) BTK fit of the low-temperature conductance curve (symbols). The use of the two-band model does not sensibly improve the fit. (b) Magnetic-field dependence of the conductance curves of the same contact. All the measurements were taken at 1.8 K. (c) Temperature dependence of the OP extracted from the single-band BTK fit of the curves shown in (a). The trend is definitely non-BCS and very similar to that shown in Fig.3b and 4b for $\Delta_1$. The dashed line is only a guide to the eye. (d) Magnetic-field dependence of the OP extracted from the fit of the curves shown in panel b. This behaviour is compatible with that predicted for type-II superconductors, and with a critical field of the order of 55 T.

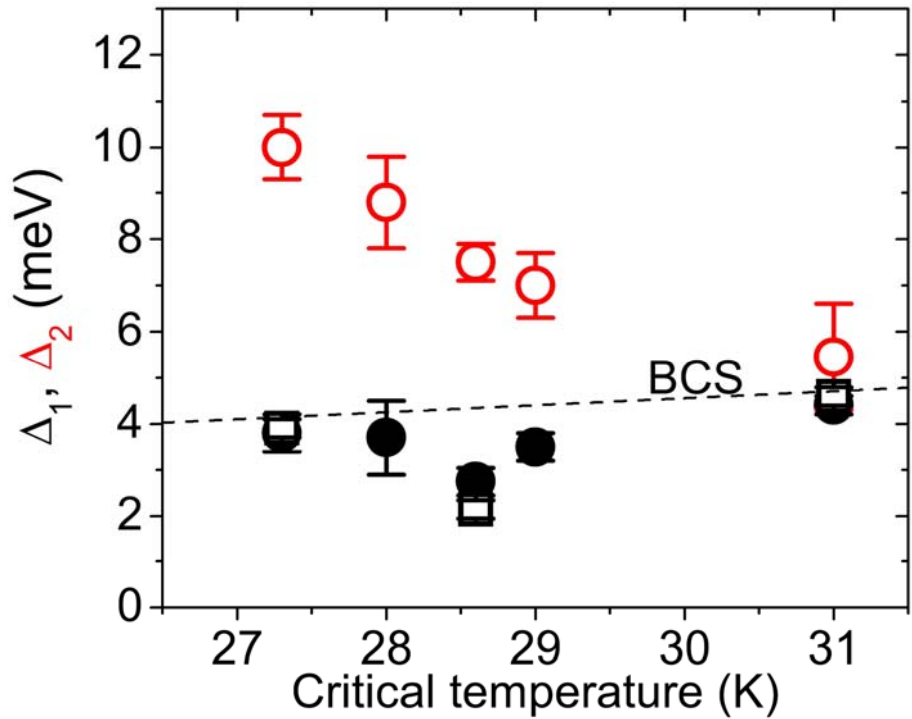

**Figure 7** (color online)
Values of $\Delta_1$ and $\Delta_2$ (solid and open circles) from the two-gap fit of the low-temperature conductance curves of different point contacts, reported as a function of the critical temperature of the contact. Gap values obtained from the single-gap fit of some of these curves are reported as well (open squares). Error bars also take into account the different normalizations. The spread in $T_c$ values is due to the fact that the sample presents homogeneous, large crystallites with various F concentrations embedded in a disordered matrix. The measurements show that $\Delta_1$ is always close or below the BCS value (solid line) and increases with $T_c$, while $\Delta_2$ is well above the BCS line and decreases on increasing $T_c$, until a single (or two, very close to each other) BCS OP is observed.

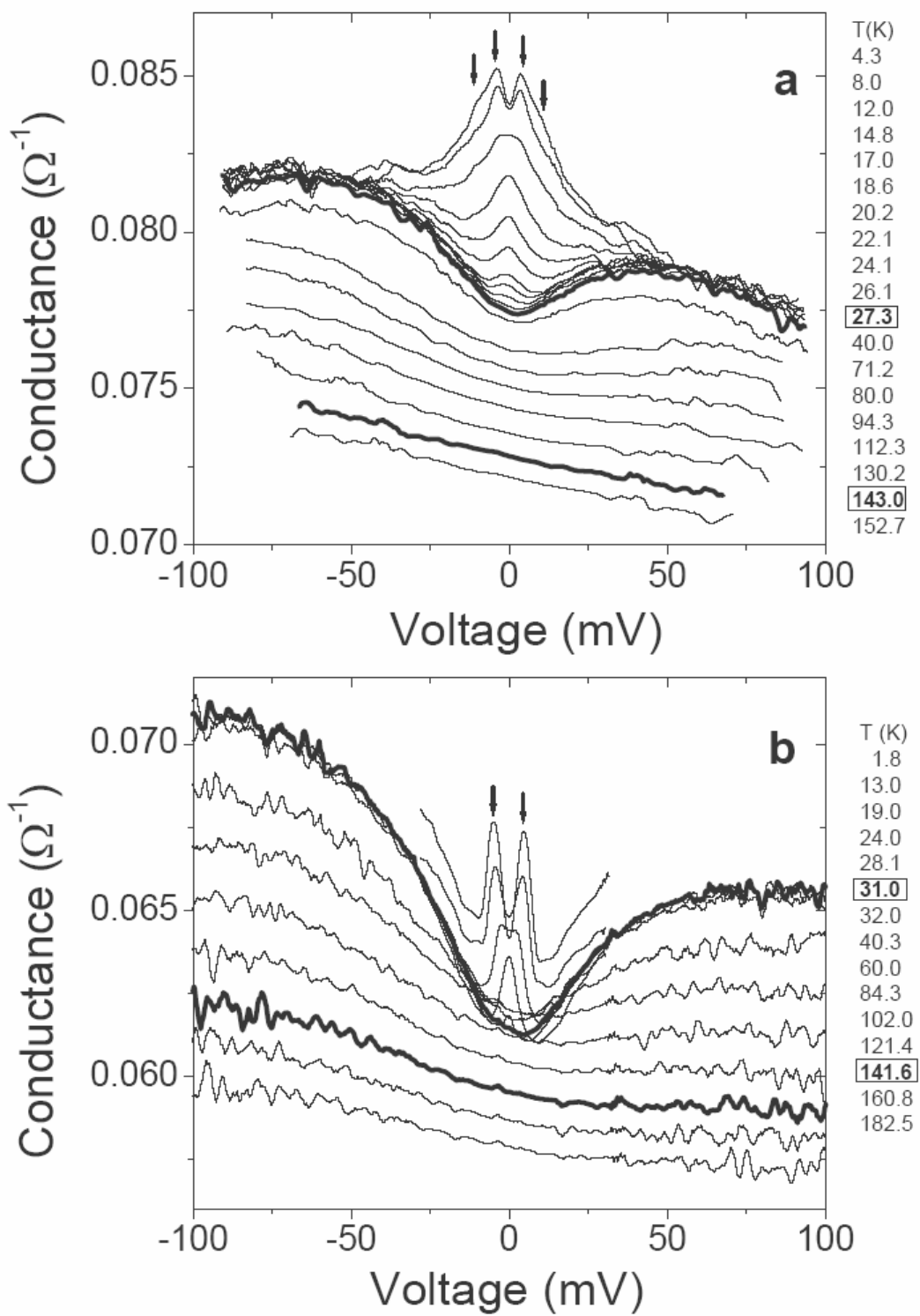

**Figure 8**

Two examples of temperature dependence of the conductance curves measured up to temperatures well above T$_c$. The first non-superconducting conductance curve (upper thick line) clearly shows a pseudogaplike feature which is progressively filled on increasing the temperature, and disappears at about 140 K (bottom thick line). The temperatures corresponding to the thick lines are highlighted in the labels.

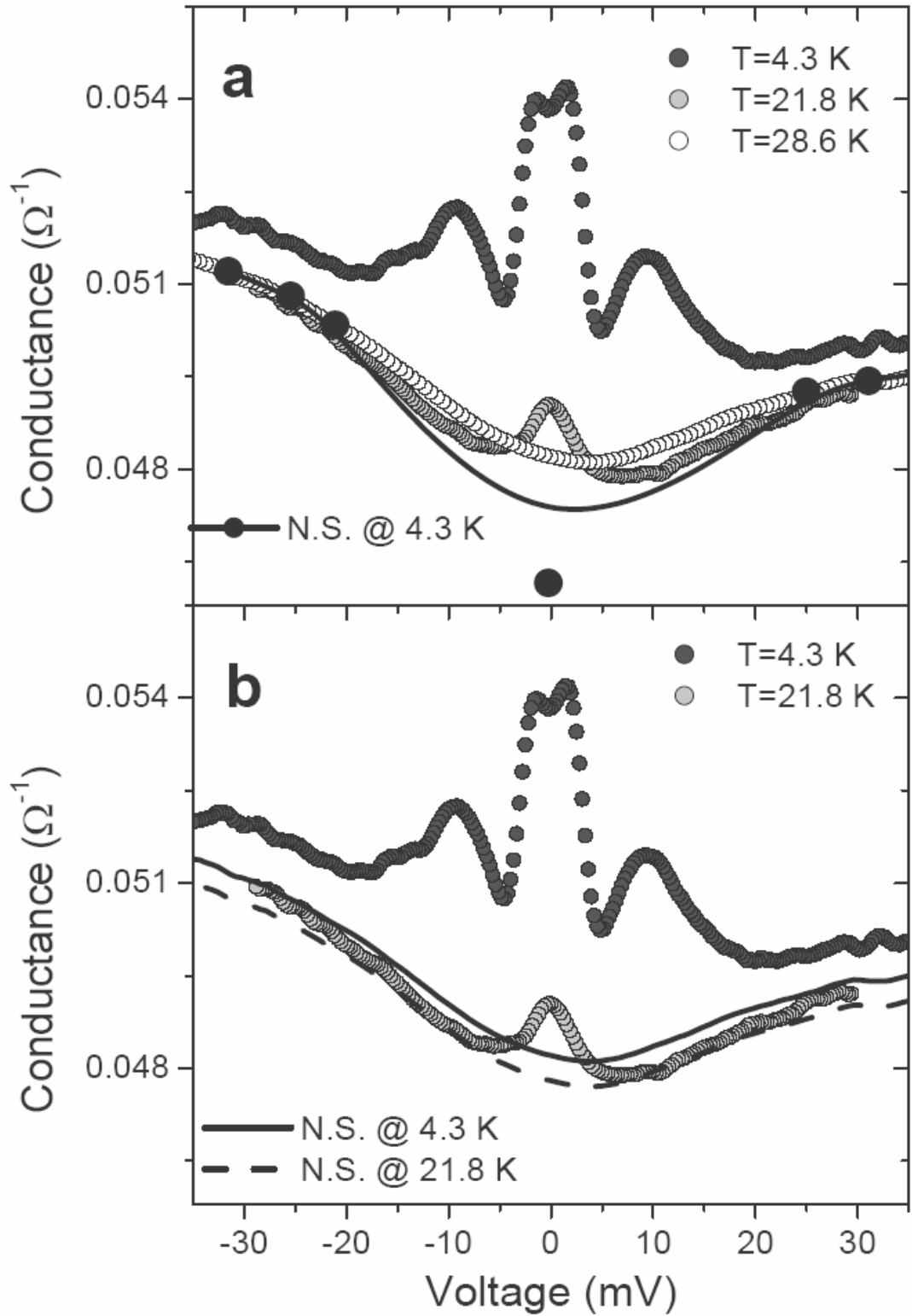

**Figure A1**
Explanation of the two alternative choices for the normalization of the conductance curves. (a) Dark grey circles represent the raw conductance curve at 4.3 K. The normal state at T=4.3 K is simulated by a "guess" curve (line) obtained by connecting the black points with a B-spline function. In this way, the calculated "normal state" at 4.3 K is deeper than the measured normal state at $T_c$ (open circles) respecting the progressive filling of the pseudogaplike feature suggested, for example, by the conductance curve at T=21.8 K (light grey circles). (b) All the curves are divided by the measured normal state at $T_c$ (solid line), translated if necessary (e.g. at T=21.8 K, dashed line).

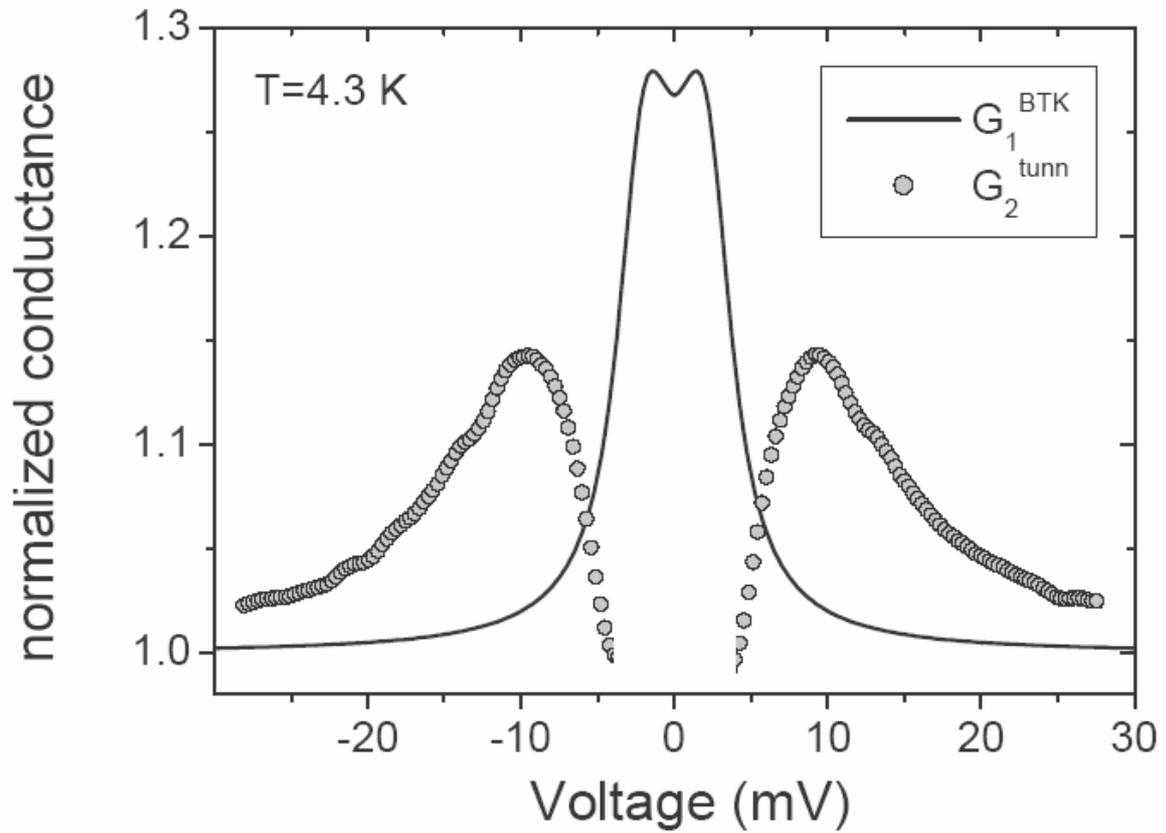

**Figure C1**
The two contributions $G_1^{BTK}$ (line) and $G_2^{tunn}$ (symbols) to the total point-contact conductance shown in Fig.3a. The sum of these contributions, divided by 2, gives the original conductance curve. $G_1^{BTK}$ is obtained by fitting the central part of the original conductance curve, disregarding the additional features at higher energy. $G_2^{tunn}$ is then obtained by subtraction, as explained in the text.

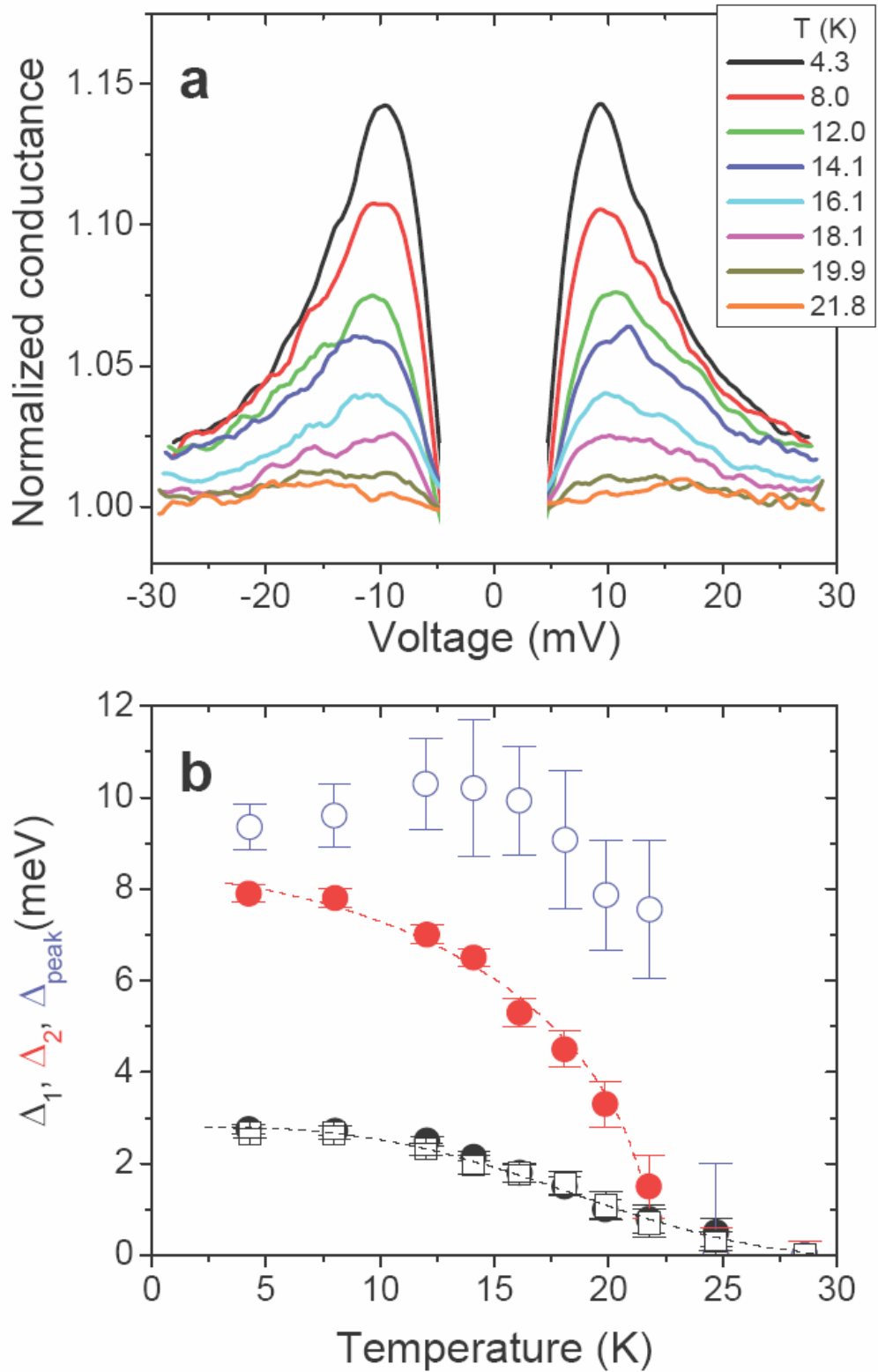

**Figure C2** (color online)
(a) Temperature dependence of $G_2^{tunn}$ obtained from the conductance curves of Fig. 3a. (b) Temperature dependence of the peak position $\Delta_{peak}$ in $G_2^{tunn}$ (open circles) and of the values of $\Delta_1$ from $G_1^{BTK}$ (open squares) compared to the result of the two-band BTK fit of the same curves (solid symbols).

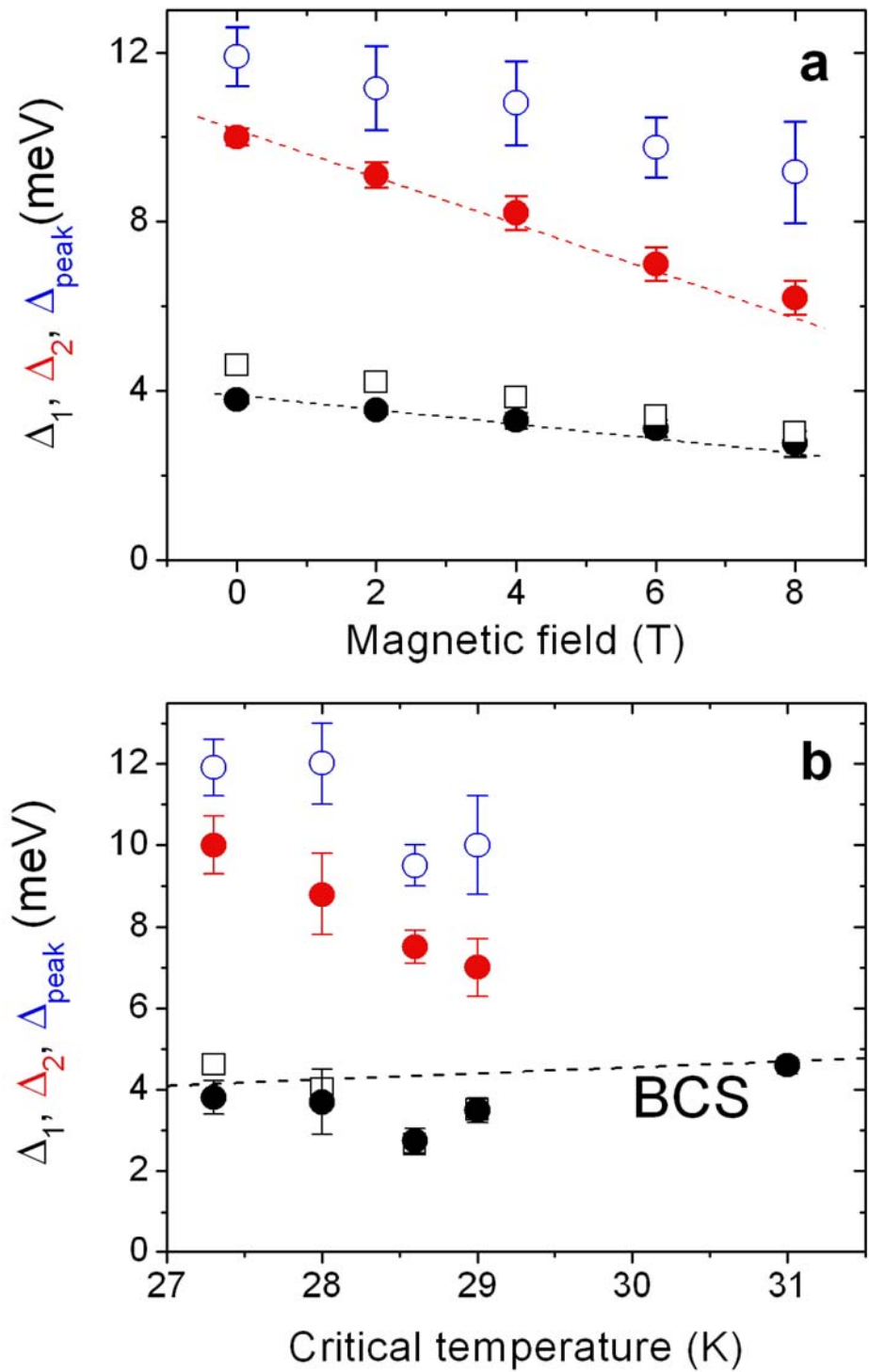

**Figure C3** (color online)
(a) Magnetic-field dependence of $\Delta_{peak}$ from $G_2^{tunn}$ (open circles) and of the values of $\Delta_1$ from $G_1^{BTK}$ (open squares) obtained from the conductance curves of Fig.5(a), compared to the results of the two-band BTK fit (solid symbols). (b) Values of $\Delta_1$ and $\Delta_{peak}$ (open squares and open circles) from the superconducting and tunnelling contributions to the low-temperature conductance curves of different point contacts, reported as a function of the local $T_c$ of the contact. The result of the two-band BTK fit is also shown (solid symbols) for comparison.